\documentclass[12pt]{article}
\usepackage{amsfonts}
\begin{document}

\def\ba{\begin{eqnarray}}
\def\ea{\end{eqnarray}}

 % \begin{titlepage}
\title{\bf The Symmetric Teleparallel Gravity}
\author{ {\bf \small Muzaffer ADAK}  \\
 {\it \small Department of Physics, Faculty of Arts and Sciences, Pamukkale University,} \\
 {\it \small  20100 Denizli-TURKEY} \\
 {\small \it e-mail: madak@pau.edu.tr} }
\vskip 1cm
  \date{ }
\maketitle

 \begin{center}
  {\small Received 18.03.2006}
 \end{center}

 \vskip 0.5cm

\begin{abstract}
We study symmetric teleparallel (STP) gravity model, in which only
spacetime non-metricity is nonzero. First we obtain STP equivalent
Einstein-Hilbert Lagrangian and give an approach for a generic
solution in terms of only metric tensor. Then we obtain a
spherically symmetric static solution to the Einstein's equation
in STP space-time and discuss the singularities. Finally, we study
a model given by a Lagrangian 4-form quadratic in non-metricity.
Thus, we seek Schwarzschild-type solutions because of its
observational success and obtain some sets of solutions. Finally,
we discuss physical relevance of the solutions. 
\end{abstract}
 %\end{titlepage}

\section{ Introduction}

    \hskip 0.6cm It can be thought that Einstein's general relativity
(GR) is one of the biggest achievements of the last century. This
work, first time, formulated  a comprehensive theory containing
gravity and matter that gave rise to a new  understanding of
universe. Some deficiencies, however, appeared in Einstein's
approach in last decades and people started to investigate whether
GR was a unique and basic theory that explains exactly the
gravitational interactions. These matters come from basically
cosmology and the quantum field theory. In the former; the
standard cosmology model based on GR and the standard model of
particle physics is inadequate for explaining the universe at
limit zones because of the existence of the bing bang singularity,
flatness and horizon problems. On the other hand, if someone wants
to achieve the quantum explanation of the space-time (or
gravitation), it is realized that GR is a {\it classical} theory.
Because of these realities and the absence of a definite quantum
gravity theory, efforts of finding an alternative gravity theory
are continued.

One of the most efficient approaches is the non-Riemannian
formulation of gravity (see \cite{adak2003} and references
therein), but little evidence for physical relevance of additional
fields. In non-Riemannian gravity models metric, co-frame and full
connection are considered as gauge potentials. The corresponding
field strengths are the non-metricity ${Q^a}_b$, the torsion $T^a$
and the curvature ${R^a}_b$. Because of the lack of experimental
results for ${Q^a}_b$ and $T^a$, in general, the non-Riemannian
gravity models are studied theoretically. Classification of the
space-time and related theories are given summarily in table
\ref{classification}.
\begin{table}[h]
 \begin{center}
   \caption{Classification of space-times} \label{classification}
  \begin{tabular}{|l||l||l|}
  \hline
  \hskip 1cm {\bf Spacetime}      & {\bf Physical Theory} & {\bf Literature} \\
  \hline \hline
  ${Q^a}_b = 0$, $T^a = 0$, ${R^a}_b = 0$    & Special Relativity    & A. Einstein (1905) \\
  \hskip 1 cm {\it Minkowski}               &                       & and many people \\
   \hline
  ${Q^a}_b = 0$, $T^a = 0$, ${R^a}_b \neq 0$ & General Relativity    & A. Einstein (1916) \\
   \hskip 0.3 cm {\it (Pseudo-)Riemannian}     &                       & and many people \\
   \hline
  ${Q^a}_b = 0$, $T^a \neq 0$, ${R^a}_b=0$ & Teleparallel Gravity & K. Hayashi \& T. Nakano (1967) \\
  \hskip 0.9cm {\it Weitzenb\"{o}ck}      &                  & and many people \cite{hayashi1967}-\cite{arcos2004}\\
   \hline
  ${Q^a}_b \neq 0$, $T^a=0$, ${R^a}_b=0$  & Symmetric Teleparallel  & J.M. Nester \& H.J.Yo (1999)  \\
   \hskip 1.2cm {\it ???????}              & Gravity               & and few people \cite{nester1999}-\cite{adak2006} \\
   \hline
  ${Q^a}_b \neq 0$, $T^a=0$, ${R^a}_b \neq 0$ & Einstein-Weyl    & H. Weyl (1919) \\
  \hskip 0.8cm {\it Riemann-Weyl}           &                    & and some people \cite{dereli1994}-\cite{hehl2005} \\
   \hline
  ${Q^a}_b = 0$, $T^a \neq 0$, ${R^a}_b \neq 0$ & Einstein-Cartan & A. Trautman (1972) \\
  \hskip 0.7cm {\it Riemann-Cartan}           &                     & and many people \cite{dereli1982}-\cite{adak2001} \\
    \hline
  ${Q^a}_b \neq 0$, $T^a \neq 0$, ${R^a}_b=0$ & \hskip 1cm  ??????? & \hskip 1.2cm  ??????? \\
   \hskip 1.2cm {\it ???????}            &             &     \\
     \hline
  ${Q^a}_b \neq 0$, $T^a \neq 0$, ${R^a}_b \neq 0$ & Einstein-Cartan-Weyl & A few \cite{dereli1987}-\cite{adak2004}\\
   \hskip 1cm {\it Non-Riemannian}    &                         &           \\
  \hline
  \end{tabular}
 \end{center}
\end{table}

As seen from the table there is nearly no work on the symmetric
teleparallel gravity (STPG). This work aims to fill this gap. Due
to the fact that curvature and torsion vanish, it is usually
asserted that this model is a gravitational field theory that is
closest possible to flat space-time. Here we adhere to the
following conventions: $\alpha$, $\beta , \;\; \cdots =
\hat{0},\hat{1},\hat{2},\hat{3}$ are holonomic or coordinate
indices and $a$, $b, \;\; \cdots = 0,1,2,3$ are anholonomic or
frame indices. Vierbein (tetrad) ${h^a}_\alpha$ and its inverse
${h^\alpha}_a$ (i.e. ${h^a}_\alpha  {h^\alpha}_b =\delta^a_b $)
give transformations between them. Abbreviations $e^{ab \cdots} =
e^a \wedge e^b \wedge \cdots$ and $(ab)=\frac{1}{2} (a+b)$ and
$[ab]=\frac{1}{2}  (a-b)$ are used.

\section{ Mathematical Preliminaries}

The triple $ \{M,g,\nabla \} $ denotes the space-time where $M$ is
a $4$-dimensional differentiable manifold, $ g $ is a
non-degenerate Lorentzian metric and $ \nabla $ is a linear
connection. $g$ can be written in terms of the co-frame $1$-forms
 \ba
      g =g_{\alpha \beta} dx^\alpha \otimes dx^\beta = \eta_{ab}e^a \otimes e^b
 \ea
where $\{ e^a \}$ orthonormal and $\{ dx^\alpha \}$  co-ordinate
co-frame $1-$forms, and $\eta_{ab}=(-,+,+,+)$ orthonormal and
$g_{\alpha \beta}$ co-ordinate components of the metric.
Orthonormal co-frame is dual to orthonormal frame $e^b(X_a)=
\imath_a e^b = \delta^b_a $. Similarly,
$dx^\beta(\partial_\alpha)= \imath_\alpha dx^\beta =
\delta^\beta_\alpha $. Here $\imath$ denotes the interior product
operator mapping any $p$-form into $(p-1)$-form. Besides, we set
space-time orientation by $\epsilon_{0123}=+1 $ or $*1 =e^{0123}$
where $*$ is the Hodge star operator mapping any $p$-form into
$(4-p)$-form. Finally, the connection is specified by a set of
connection $1$-forms $\{ {\Lambda^a}_b \}$. In the gauge approach
to gravity, $\eta_{ab} , \quad e^a , \quad {\Lambda^a}_b$ are
interpreted as the generalized gauge potentials, and the
corresponding field strengths; the non-metricity $1$-forms,
torsion $2$-forms and curvature $2$-forms are defined through the
Cartan structure equations; table \ref{fieldstrenghts}.

\begin{table}[h]
 \begin{center}
 \caption{Gauge potentials and field strengths}\label{fieldstrenghts}
  \begin{tabular}{|ll||ll|}
   \hline
      & {\bf Gauge Potential} & {\bf Field Strength} & \\
   \hline \hline
   $\eta_{ab}$ & {\it o.n. metric} & $ Q_{ab} := -\frac{1}{2} D\eta_{ab} = \frac{1}{2}(\Lambda_{ab} +\Lambda_{ba})$ & {\it Nonmetricity $1-$form} \\
   \hline
   $e^a$ & {\it o.n. co-frame} & $T^a := De^a = de^a + {\Lambda^a}_b \wedge e^b$ & {\it Torsion $2-$form}\\
   \hline
   ${\Lambda^a}_b$ & {\it Full connection } & ${R^a}_b := D{\Lambda^a}_b := d{\Lambda^a}_b +{\Lambda^a}_c \wedge {\Lambda^c}_b$ & {\it Curvature $2-$form}\\
   \hline
  \end{tabular}
 \end{center}
\end{table}

 \noindent Here $d$ is the exterior derivative mapping any $p$-form into
$(p+1)$-form. These field strengths satisfy the Bianchi
identities; table \ref{bianchiidentities}.

\begin{table}[h]
 \begin{center}
 \caption{Bianchi identities}\label{bianchiidentities}
  \begin{tabular}{|l|l|}
    \hline
    $ DQ_{ab} = \frac{1}{2}(R_{ab} +R_{ba})$ & {\it Zeroth} \\
    \hline
    $ DT^a = {R^a}_b \wedge e^b$ & {\it First} \\
    \hline
    $ D{R^a}_b = 0 $ & {\it Second}  \\
    \hline
  \end{tabular}
 \end{center}
\end{table}

 \noindent We also need the identities \footnote{Since
$Q^{ab}=\frac{1}{2}D\eta^{ab} \neq 0$ special attention in
lowering and raising index in front of covariant exterior
derivative.}
 \ba
    D*e_a &=& - Q \wedge *e_a + T^b \wedge *e_{ab} \\
    D*e_{ab} &=& - Q \wedge *e_{ab} + T^c \wedge *e_{abc} \label{Deab}\\
    D*e_{abc} &=& - Q \wedge *e_{abc} + T^d \wedge *e_{abcd} \\
    D*e_{abcd} &=& - Q \wedge *e_{abcd}
 \ea
where with $Q={\Lambda^a}_a={Q^a}_a$ Weyl $1$-form. The full
connection $1$-forms are decomposed uniquely as follows
\cite{dereli1987}-\cite{hehl1995}:
 \ba
  {\Lambda^a}_b = {\omega^a}_b + {K^a}_b
          + {q^a}_b + {Q^a}_b \;  \label{connec}
 \ea
where $ {\omega^a}_b $ are the Levi-Civita connection $1$-forms
 \ba
      {\omega^a}_b \wedge e^b = - de^a  \quad \quad \mbox{or} \quad \quad
  2\omega_{ab} = -\imath_a (de_b) + \imath_b (de_a) +
  [\imath_a\imath_{b}(de_c)]e^c \; ,
 \ea
$ {K^a}_b $ are the contortion $1$-forms
 \ba
     {K^a}_b \wedge e^b = T^a \quad \quad \mbox{or} \quad \quad
  2K_{ab} = \imath_aT_b- \imath_bT_a - (\imath_a\imath_{b}T_c)e^c
  \; ,
 \ea
and $ {q^a}_b $ anti-symmetric tensor $1$-forms
 \ba
     q_{ab} = -(\imath_a Q_{bc}) \wedge e^c
        + (\imath_b Q_{ac}) \wedge e^c \; .
 \ea
In this decomposition the symmetric part
 \ba
    \Lambda_{(ab)} = Q_{ab}
 \ea
while the anti-symmetric part
 \ba
  \Lambda_{[ab]} = \omega_{ab} + K_{ab} + q_{ab} \; .
 \ea

In gravity models it is complicated to keep all the components of
${Q^a}_b$. Therefore, people sometimes deal only with certain
irreducible parts of that. To obtain the irreducible
decompositions of non-metricity invariant under the Lorentz group,
firstly we write
 \ba
   Q_{ab} = \underbrace{\overline{Q}_{ab}}_{trace-free \;\; part} + \underbrace{\frac{1}{4} \eta_{ab}Q}_{trace \;\; part}
 \ea
where Weyl $1$-form $Q={Q^a}_a$ and $ \eta^{ab}\overline{Q}_{ab} =
0$. Now we sum up the components
 \ba
     Q_{ab} = { }^{(1)}Q_{ab} + { }^{(2)}Q_{ab} + { }^{(3)}Q_{ab} + { }^{(4)}Q_{ab}
 \ea
in terms of
 \ba
    { }^{(2)}Q_{ab} &=& \frac{1}{3} {}*(e_a \wedge \Omega_b +e_b \wedge \Omega_a) \; ,  \\
    { }^{(3)}Q_{ab} &=& \frac{2}{9}( \Lambda_a e_b +\Lambda_b e_a -\frac{1}{2} \eta_{ab} \Lambda ) \; , \\
    { }^{(4)}Q_{ab} &=&  \frac{1}{4} \eta_{ab} Q  \; ,  \\
    { }^{(1)}Q_{ab}  &=& Q_{ab}- { }^{(2)}Q_{ab} - { }^{(3)}Q_{ab} - { }^{(4)}Q_{ab}
 \ea
where
 \ba
    \Lambda_b &:=& \imath_a { \overline{Q}^a}_b \;\; , \;\;\;\;\;\;\;\;
                   \Lambda := \Lambda_a e^a,    \nonumber  \\
    \Theta_b &:=& {}*(\overline{Q}_{ab} \wedge e^a) \;\; , \;\;\;
    \Theta := e^b \wedge \Theta_b \;\; , \;\;\;
    \Omega_a := \Theta_a -\frac{1}{3}\imath_a\Theta \; .
 \ea
The components have properties
 \ba
   \eta_{ab} { }^{(1)}Q^{ab} = \eta_{ab} { }^{(2)}Q^{ab}
     =\eta_{ab} { }^{(3)}Q^{ab} = 0 \; ,  \nonumber \\
 \imath_a { }^{(1)}Q^{ab} = \imath_a { }^{(2)}Q^{ab} =0 \;, \nonumber \\
  e_a \wedge { }^{(1)}Q^{ab} =0 \; ,  \nonumber \\
  \imath_{(a}{ }^{(2)}Q_{bc)} =0 \; .
 \ea
Thus the components are orthogonal in the following sense
 \ba
    { }^{(i)}Q^{ab} \wedge \; * { }^{(j)}Q_{ab} = \delta^{ij}
    N_{ij} \quad \quad (\mbox{\small no summation over {\it ij}})
 \ea
where $\delta^{ij}$ is the Kronecker symbol and $N_{ij}$ any
$4$-form. Then
 \ba
 { }^{(1)}Q^{ab} \wedge \; * { }^{(1)}Q_{ab} &=& Q^{ab} \wedge \;* Q_{ab}
        - { }^{(2)}Q^{ab} \wedge \; * { }^{(2)}Q_{ab} - { }^{(3)}Q^{ab} \wedge \; * { }^{(3)}Q_{ab} \nonumber \\
     & &- { }^{(4)}Q^{ab} \wedge \; * { }^{(4)}Q_{ab} \; , \label{Q1Q1} \\
 { }^{(2)}Q^{ab} \wedge \; * { }^{(2)}Q_{ab} &=& \frac{2}{3} (Q_{ac} \wedge e^a) \wedge * (Q^{bc} \wedge e_b)
 - \frac{2}{9} (\imath^a Q_{ac}) (\imath_b Q^{bc}) *1 - \frac{2}{9} Q \wedge * Q \nonumber \\
    & & + \frac{4}{9} (\imath_a Q) (\imath_b Q^{ab}) *1 \; , \\
 { }^{(3)}Q^{ab} \wedge \; * { }^{(3)}Q_{ab} &=& \frac{4}{9} (\imath^a
 Q_{ac}) (\imath_b Q^{bc}) *1 + \frac{1}{36} Q \wedge * Q
 - \frac{2}{9} (\imath_a Q) (\imath_b Q^{ab}) *1 \; , \\
 { }^{(4)}Q^{ab} \wedge \; * { }^{(4)}Q_{ab} &=& \frac{1}{4}   Q \wedge * Q \;
 . \label{Q4Q4}
 \ea

 \section{ Symmetric Teleparallel Gravity}

STP space-time is defined as ${Q^a}_b \neq 0 \;  , \; {T^a} = 0 \;
, \; {R^a}_b = 0 $. One generic solution to that is obtained in
the coordinate frame;  ${\Lambda^\alpha}_\beta =0$, the so-called
{\it natural or inertial gauge}:
  \ba
    {R^\alpha}_\beta = d{\Lambda^\alpha}_\beta +
           {\Lambda^\alpha}_\gamma \wedge {\Lambda^\gamma}_\beta =0 \; ,\\
  T^\alpha = d(dx^\alpha) +{\Lambda^\alpha}_\beta \wedge dx^\beta  =0 \; ,\\
  Q_{\alpha \beta} =- \frac{1}{2}Dg_{\alpha \beta} = - \frac{1}{2}dg_{\alpha \beta} \neq
  0 \; .
  \ea
After a frame transformation via vierbein  $e^a = {h^a}_\alpha
dx^\alpha$ and ${\Lambda^a}_b = {h^a}_\alpha
{\Lambda^\alpha}_\beta {h^\beta}_b + {h^a}_\alpha d {h^\alpha}_b$
we obtain the field strengths in orthonormal components
 \ba
    {R^a}_b = {h^a}_\alpha {R^\alpha}_\beta {h^\beta}_b =0 \; ,\\
  T^a = {h^a}_\alpha T^\alpha  =0 \; ,\\
  Q_{ab} = Q_{\alpha \beta} {h^\alpha}_a {h^\beta}_b  \neq 0 \; .
 \ea

 \subsection{The Einstein-Hilbert Lagrangian in STPG}

The orthonormal teleparallel representation of the Einstein's
theory is interesting and useful. Therefore, we derive the
Einstein-Hilbert Lagrangian $4$-form. Firstly we use the
decomposition of the full connection (\ref{connec}), with ${K^a}_b
=0$,
 \ba
   {\Lambda^a}_b = {\omega^a}_b + {\Omega^a}_b \quad \mbox{where}
   \quad {\Omega^a}_b = {Q^a}_b + {q^a}_b \; .
 \ea
By substituting that into ${R^a}_b (\Lambda)$ we decompose the
non-Riemannian curvature as the follows:
 \ba
  {R^a}_b (\Lambda) &=& d{\Lambda^a}_b +{\Lambda^a}_c \wedge  {\Lambda^c}_b \nonumber \\
                    &=& {R^a}_b (\omega) +D(\omega){\Omega^a}_b +{\Omega^a}_c \wedge
                    {\Omega^c}_b \label{RabL}
 \ea
where ${R^a}_b (\omega)$ is the Riemannian curvature $2$-form and
$D(\omega)$ is the covariant exterior derivative with respect to
the Levi-Civita connection. To set ${R^a}_b (\Lambda)=0$ for STP
space-time yields the Einstein-Hilbert Lagrangian $4$-form
  \ba
    L_{EH} &=& {R^a}_b (\omega) \wedge *{e_a}^b \nonumber \\
           &=& - [ D(\omega){\Omega^a}_b ] \wedge *{e_a}^b
                 - {\Omega^a}_c \wedge  {\Omega^c}_b \wedge *{e_a}^b
  \ea
Here after using the equality
 \ba
     d ( {\Omega^a}_b  \wedge *{e_a}^b) = [ D(\omega){\Omega^a}_b ] \wedge *{e_a}^b
         - {\Omega^a}_b  \wedge [D(\omega)*{e_a}^b]
 \ea
we discard the exact form and we notice that $D(\omega)*{e_a}^b =0
$ because $T^a$ and ${Q^a}_b$ vanish for  ${\omega^a}_b$ (see
eq.(\ref{Deab})). Thus
 \ba
      L_{EH} &=& \frac{1}{2\kappa} {\Omega^a}_c \wedge  {\Omega^c}_b \wedge *{e_a}^b  \nonumber \\
          &=& \frac{1}{2\kappa} (Q^{ac}+q^{ac}) \wedge (Q_{cb}+q_{cb})  \wedge *{e_a}^b \nonumber \\
      &=& \frac{1}{2\kappa} (Q^{ac} \wedge Q_{cb} +q^{ac} \wedge q_{cb}) \wedge *{e_a}^b \nonumber \\
  &=& \frac{1}{2\kappa} [ -Q_{ab} \wedge *Q^{ab} + 2 (Q_{ac} \wedge e^a) \wedge *(Q^{bc} \wedge e_b)
    -Q \wedge *Q + 2 (\imath_b Q)(\imath_a Q^{ab})*1]
 \ea
where $\kappa$ is gravitational coupling constant. In
two-dimension we note $ q^{ac} \wedge q_{cb} \wedge *{e_a}^b =0$.

 \subsection{A Symmetric Teleparallel Solution to the Einstein Equation}

Now we give a brief outline of GR. GR is written in (pseudo-)
Riemannian spacetime in which torsion and non-metricity are both
zero, i.e., connection is Levi-Civita. Einstein equation can be
written in the following form
  \ba
     G_a := -\frac{1}{2}R^{bc}(\omega) \wedge { }*e_{abc} = \kappa \tau_a \label{einsteineqn}
  \ea
or alternatively
 \ba
     { }*G_a := \mbox{(Ric)}_a - \frac{1}{2} \mathcal{R} e_a = \kappa { }*\tau_a
 \ea
where $G_a$ is Einstein tensor $3$-form, $R^{ab}(\omega)$ is
Riemannian curvature $2$-form, $\mbox{(Ric)}^a = \imath^b {R^a}_b
(\omega)$ is Ricci curvature $1$-form, $ \mathcal{R} =\imath_a
\mbox{(Ric)}^a $ is scalar curvature, $\tau_a $ is the
energy-momentum $3$-form. For the symmetric teleparallel
equivalent of Einstein equation we use the the decomposition of
the non-Riemannian curvature $2$-form (\ref{RabL}) and set
${R^a}_b(\Lambda ) = 0$. Thus we obtain the symmetric teleparallel
equivalent of (\ref{einsteineqn})
 \ba
   G_a :=\frac{1}{2} [ D(\omega )q^{bc} + {q^b}_k \wedge q^{kc} + {Q^b}_k \wedge Q^{kc} ]
       \wedge { }*e_{abc}  =  \kappa \tau_a \; . \label{stpegr}
 \ea

We now proceed the attempt for finding a solution to the STPG
model. As usual in the study of exact solutions, we have two
steps. The first one is to choose the convenient local coordinates
and make corresponding ansatz for the dynamical fields. The second
step concerns providing the invariants of the resulting geometry.
While the choice of an ansatz helps to solve the field equations
easily, the invariant description provides the correct
understanding of the physical contents of a solution.

Since metric and connection are independent quantities in
non-Riemannian spacetimes, we have to predict separately
appropriate candidates for them. Therefore we first write a line
element in order to determine the metric. We naturally start
dealing with the case of spherical symmetry for realistic
simplicity,
 \ba
     g=-f^2 dt^2 + g^2dr^2 + r^2d\theta^2 +r^2\sin^2\theta
     d\varphi^2
 \ea
where $f=f(r)$ and $g=g(r)$. A convenient choice for a tetrad
reads
 \ba
      e^0 = fdt , \quad e^1= gdr , \quad e^2= rd\theta ,
      \quad e^3 = r\sin\theta d\varphi \; . \label{coframe1}
 \ea
In addition, for the non-Riemannian connection we choose
 \ba
     \Lambda_{12} &=& -\Lambda_{21}= - \frac{1}{r}e^2 , \quad
     \Lambda_{13}=-\Lambda_{31}= -  \frac{1}{r}e^3  , \quad
     \Lambda_{23}=-\Lambda_{32}= - \frac{\cot\theta}{r}e^3 , \nonumber \\
     \Lambda_{00} &=& \frac{f'}{fg}e^1 , \quad
     \Lambda_{11} = \frac{1}{r}(1-\frac{1}{g})e^1 , \quad
     \Lambda_{22} = \frac{1}{r}(1-\frac{1}{g})e^1 , \nonumber \\
     \Lambda_{33} &=& \frac{1}{r}(1-\frac{1}{g})e^1 , \quad \mbox{others}=0  \label{connect1}
 \ea
where prime denotes derivative with respect to $r$. These gauge
configurations (\ref{coframe1}) and (\ref{connect1}) satisfy the
constraint equations ${R^a}_b(\Lambda )=0 \; , \quad T^a(\Lambda
)=0$. One can certainly perform a locally Lorentz transformation
 \ba
      e^a \rightarrow {L^a}_b e^b \quad , \quad
      {\Lambda^a}_b \rightarrow {L^a}_c {\Lambda^c}_d {{L^{-1}}^d}_b
      +{L^a}_c d{{L^{-1}}^c}_b
 \ea
which yields the Minkowski gauge ${\Lambda^a}_b =0$. This may mean
that we propose a set of connection components in a special frame
and coordinate which seems contrary to the spirit of relativity
theory. However in physically natural situations we can choose a
reference and coordinate system at our best convenience.

We deduce from equations (\ref{coframe1})-(\ref{connect1})
 \ba
    \omega_{01} &=& -\frac{f'}{fg}e^0 , \quad \omega_{12}=- \frac{1}{rg} e^2 , \quad
     \omega_{13}=- \frac{1}{rg} e^3 , \quad
     \omega_{23}= -\frac{\cot\theta}{r} e^3 \; ,\nonumber \\
  Q_{00} &=& \frac{f'}{fg}e^1 , \quad Q_{11} = \frac{1}{r}(1-\frac{1}{g})e^1 , \quad
     Q_{22} = \frac{1}{r}(1-\frac{1}{g})e^1 , \quad
     Q_{33} = \frac{1}{r}(1-\frac{1}{g})e^1  \; , \nonumber \\
   q_{01} &=& \frac{f'}{fg}e^0 , \quad q_{12} = \frac{1}{r}(\frac{1}{g}-1)e^2 , \quad
     q_{13} =\frac{1}{r}(\frac{1}{g}-1) e^3  , \quad   \mbox{others}=0 \; .
     \label{oQq1}
 \ea
When we put (\ref{oQq1}) into (\ref{stpegr}) we obtain, with
$\tau_a = 0 $
 \ba
     \left( dq^{bc} + 2{\omega^b}_f \wedge q^{fc}
      + {q^b}_f \wedge q^{fc} \right) \wedge { }*e_{abc} =0
 \ea
whose components read explicitly
 \ba
     Zeroth \;\; component \quad \quad \quad \quad  \quad \quad \;\;
                                 \left[ \frac{2(g^{-1})'}{rg} - \frac{g^2 -1}{r^2g^2} \right]
                                        e^{123} = 0 \label{zeroth}\\
     First \;\; component \quad \quad \quad \quad  \quad \quad \quad
                                        -\left[ \frac{2f'}{rfg^2} - \frac{g^2 -1}{r^2g^2} \right]
                                        e^{023} = 0 \label{first}\\
     Second \;\; component \quad \quad \;\; \left[ \frac{(f'g^{-1})'}{fg}
                                    + \frac{f'}{rfg^2} + \frac{(g^{-1})'}{rg} \right]
                                       e^{013} = 0 \label{second}\\
     Third \;\; component \quad \quad - \left[ \frac{(f'g^{-1})'}{FG} + \frac{f'}{rfg^2} + \frac{(g^{-1})'}{rg} \right]
                                        e^{012} =  0 \; . \label{third}
 \ea
Then from (\ref{zeroth}) and (\ref{first})
 \ba
       g(r) &=& 1/f(r)
 \ea
and from (\ref{second}) and (\ref{third})
 \ba
      f^2(r)&=& 1- \frac{C}{r}
 \ea
where $C$ is a constant.

In order to have a correct understanding of the resulting
solution, we need to construct invariants of the {\it Riemannian}
curvature and the non-metricity. Although the total curvature is
identically zero in the teleparallel gravity, the Riemannian
curvature of the Levi-Civita connection is nontrivial:
 \ba
   R^{01}(\omega ) =\frac{(f'g^{-1})'}{fg} e^{10} \quad , \quad
   R^{02}(\omega ) =\frac{f'}{rfg^2} e^{20} \quad , \quad
   R^{03}(\omega ) =\frac{f'}{rfg^2} e^{30} \quad , \nonumber \\
   R^{12}(\omega ) =\frac{(g^{-1})'}{rg} e^{21} \quad , \quad
   R^{13}(\omega ) =\frac{(g^{-1})'}{rg} e^{31} \quad , \quad
   R^{23}(\omega ) =\frac{1}{r^2}(1-\frac{1}{g^2}) e^{32} \; .
 \ea
Thus the quadratic invariant of the Riemannian curvature reads
 \ba
  R_{ab}(\omega ) \wedge { }*R^{ab}(\omega ) &=&
                         \left\{ 2  \left[ \frac{(f'g^{-1})'}{fg}
                         \right]^2 \right.
                     + 4 \left( \frac{f'}{rfg^2} \right)^2
                         + 4 \left[ \frac{(g^{-1})'}{rg} \right]^2
                     +  \left. 2 \left[ \frac{1}{r^2} \left( 1-\frac{1}{g^2} \right) \right]^2 \right\} {
                     }*1 \nonumber \\
  &=& \frac{6C^2}{r^6} { }*1 \label{rinvar}
 \ea
and the spacetime geometry is naturally characterized by the
quadratic invariant of the nonmetricity
 \ba
     Q_{ab}\wedge { }*Q^{ab} &=&
       \left\{ \left( \frac{f'}{fg} \right)^2
          + 3 \left[ \frac{1}{r} \left( 1-\frac{1}{g} \right) \right]^2
          \right\} { }*1 \nonumber \\
       &=& \left\{ \frac{C^2}{4r^3(r-C)} - \frac{3C}{r^3} + \frac{6}{r^2}
        \left[ 1- \left( 1- \frac{C}{r}\right)^{1/2} \right] \right\} { }*1 \; .
        \label{qinvar}
 \ea
These two quadratic invariants provide the sufficient tools for
understanding the contents of the classical solutions. Important
observation is that the Riemannian curvature invariant
(\ref{rinvar}) is singular at $r=0$, but regular at the zero
($r=C$) of the metric function $f(r)$, which means that we have a
horizon here. The resulting geometry then describes the well known
Schwarzschild black hole at $r=0$ with the horizon at $r=C$. Since
we are dealing with symmetric teleparallel gravity, it is
necessary also to analyze the behavior of nonmetricity. As seen
from (\ref{qinvar}), the nonmetricity invariant diverges not only
at the origin $r=0$, but also at the Schwarzschild horizon $r=C$.
The horizon is a regular surface from the viewpoint of the
Riemannian geometry, but it is singular from the viewpoint of
symmetric teleparallel gravity.

\section{ Lagrange Formulation of STPG}

We formulate STPG in terms of a Lagrangian 4-form
 \ba
   \mathcal{L} = L + \lambda_a \wedge T^a + {R^a}_b \wedge  {\rho_a}^b \label{lagranj1}
 \ea
where ${\rho_a}^b$ and $\lambda_a$ are the Lagrange multiplier
$2$-forms giving the constraints
 \ba
   {R^a}_b = 0 \quad , \quad \quad T^a = 0 \; .
 \ea
$\mathcal{L}$ changes by a closed form under the transformations
 \ba
      \lambda_a & \rightarrow & \lambda_a + D\mu_a \; , \label{lama}\\
   {\rho_a}^b   & \rightarrow &{\rho_a}^b + D{\xi_a}^b - \mu_a \wedge
   e^b \label{rhoab}
 \ea
of the Lagrange multiplier fields. Here $\mu_a$ and ${\xi_a}^b$
are arbitrary $1$-forms. To show this invariance we use the
Bianchi identities and discard exact forms. Consequently the field
equations derived from the Lagrangian $4$-form (\ref{lagranj1})
will determine the Lagrange multipliers only up to above
transformations. The gravitational field equations are derived
from (\ref{lagranj1}) by independent variations with respect to
the connection $ \{ {\Lambda^a}_b \} $ and the ortohonormal
co-frame $\{ e^a \}$ $1$-forms, respectively:
 \ba
     \lambda_a \wedge e^b + D{\rho_a}^b = - {\Sigma_a}^b  \; , \label{dLambda1}  \\
  D\lambda_a = - \tau_a  \label{dcoframe1}
 \ea
where ${\Sigma_a}^b = \frac{\partial L}{\partial {\Lambda^a}_b}$
and $\tau_a = \frac{\partial L}{\partial e^a}$. In principle the
first field equation (\ref{dLambda1}) is used to solve for the
Lagrange multipliers $\lambda_a$ and ${\rho_a}^b$ and the second
field equation (\ref{dcoframe1}) governs the dynamics of the
gravitational fields. Here the first equation, however, has $64$
and the second one has $16$ independent components, thus giving
the total number of independent equations $80$. On the other hand,
there are totally $120$ unknowns: $24$ for $\lambda_a$ plus $96$
for ${\rho_a}^b$. But we note that the left-hand side of
(\ref{dLambda1}) is invariant under the transformations
(\ref{lama})-(\ref{rhoab}) and consequently it is sufficient to
determine the gauge invariant piece of the Lagrange multipliers,
namely $\lambda_a \wedge e^b + D{\rho_a}^b$, in terms of
${\Sigma_a}^b$. One can consult Ref.\cite{vasilic2000} for further
discussions on gauge symmetries of Lagrange multipliers. It is
important to notice $D \lambda_a$ rather than the Lagrange
multipliers themselves couple to the second field equations
(\ref{dcoframe1}). As a result we must calculate $D\lambda_a$
directly and we can manage that by taking the set of covariant
exterior derivative of (\ref{dLambda1}):
 \ba
     D\lambda_a \wedge e^b = - D {\Sigma_a}^b \; .  \label{Dlambda}
 \ea
Here we used the constraints
 \ba
     De^b &=& T^b =0 ,\\
 D^2{\rho_a}^b &=& D(D{\rho_a}^b) = {R^b}_c \wedge {\rho_a}^c - {R^c}_a \wedge
 {\rho_c}^b =0 \label{D2Sab}
 \ea
where the covariant exterior derivative of a $(1,1)$-type tensor
is
 \ba
    D{\rho_a}^b = d {\rho_a}^b + {\Lambda^b}_c \wedge {\rho_a}^c
         - {\Lambda^c}_a \wedge {\rho_c}^b \; .
 \ea
The result (\ref{Dlambda}) is unique because $D\lambda_a
\rightarrow D\lambda_a$ under (\ref{lama}).  Thus we arrive at the
field equation
 \ba
     D{\Sigma_a}^b -  \tau_a \wedge e^b =0 \; . \label{fieldeqn}
 \ea

Now we write down the following Lagrangian 4-form which is the
most general quadratic expression in the non-metricity tensor
\cite{obukhov1997}:
 \ba
   L = \frac{1}{2\kappa} \left[ k_0 {R^a}_b \wedge *{e_a}^b
    + \sum_{I=1}^4 k_I { }^{(I)}Q_{ab} \wedge * { }^{(I)}Q^{ab}
    + k_5 \left( { }^{(3)}Q_{ab} \wedge e^b \right) \wedge * \left( { }^{(4)}Q^{ac} \wedge e_c
    \right) \right] \; . \label{Lagrange}
   \ea
Here $k_0, k_1, k_2, k_3, k_4, k_5$ are dimensionless coupling
constants and $\kappa = \frac{8\pi G}{c^3}$, with $G$ the Newton's
gravitational constant. Inserting (\ref{Q1Q1})-(\ref{Q4Q4}) into
(\ref{Lagrange}) we find
 \ba
   L = \frac{1}{2\kappa} \left[ k_0 {R^a}_b \wedge *{e_a}^b
    + c_1 Q_{ab} \wedge * Q^{ab}
      +c_2 (Q_{ac} \wedge e^a) \wedge * (Q^{bc} \wedge e_b)  \right. \nonumber \\
      \left. + c_3 (\imath_a Q^{ac}) (\imath^b Q_{bc}) *1
      +c_4 Q \wedge * Q
       +c_5 (\imath_a Q) (\imath_b Q^{ab}) *1 \right] \label{lagranj}
   \ea
where the new coefficients are the following combinations of the
original coupling constants:
 \ba
 c_1 &=& k_1 ,\nonumber \\
 c_2 &=& - \frac{2}{3} k_1 + \frac{2}{3} k_2 \; ,\nonumber \\
 c_3 &=& - \frac{2}{9} k_1 - \frac{2}{9} k_2 + \frac{4}{9} k_3 \;, \nonumber \\
 c_4 &=& - \frac{1}{18} k_1 - \frac{2}{9} k_2 + \frac{1}{36} k_3 + \frac{1}{4} k_4 + \frac{1}{16} k_5 \;, \nonumber \\
 c_5 &=& - \frac{2}{9} k_1 + \frac{4}{9} k_2 - \frac{2}{9} k_3 - \frac{1}{4} k_5 \; .
 \ea
We obtain the variational field equations from (\ref{lagranj})
 \ba
   {\Sigma_a}^b = \sum_{i=0}^{5} c_i \; { }^i{\Sigma_a}^b \quad ,
   \quad \quad \quad \quad
  \tau_a = \sum_{i=0}^{5} c_i \; { }^i\tau_a
 \ea
where
 \ba
   { }^0{\Sigma_a}^b  &=& 2Q^{bc} \wedge *e_{ac} - Q \wedge *{e_a}^b + T_c \wedge *{e_a}^{bc} \\
     { }^1{\Sigma_a}^b  &=& *({Q_a}^b + {Q^b}_a) \\
  { }^2{\Sigma_a}^b  &=& e_a \wedge *(Q^{bc} \wedge e_c) + e^b \wedge *(Q_{ac} \wedge e^c) \\
   { }^3{\Sigma_a}^b  &=& \imath^c Q_{ac} *e^b + \imath_c Q^{bc} *e_a \\
    { }^4{\Sigma_a}^b  &=& 2 \delta^b_a { }*Q \\
    { }^5{\Sigma_a}^b  &=& \frac{1}{2} (\imath^b Q) *e_a +  \frac{1}{2} (\imath_a Q) *e^b
                          + \delta^b_a  (\imath_c Q^{cd}) *e_d \\
    { }^0\tau_a  &=&  {R^b}_c \wedge *{e_{ab}}^c \\
    { }^1\tau_a  &=&  - (\imath_a Q^{bc}) \wedge *Q_{bc}
                              - Q^{bc} \wedge (\imath_a *Q_{bc}) \\
    { }^2\tau_a  &=& -Q_{ab} \wedge *(Q^{bc} \wedge e_c)
                            - (\imath_a Q^{cd}) e_c \wedge *(Q_{bd} \wedge e^b )
                             + (Q_{bd} \wedge e^b) \wedge *(Q^{cd} \wedge e_{ca}) \\
    { }^3\tau_a  &=& -2 (\imath_a Q^{bd}) (\imath^c Q_{cd}) *e_b
                             + (\imath_b Q^{bd}) (\imath^c Q_{cd}) *e_a \\
    { }^4\tau_a  &=&  -(\imath_a Q) *Q - Q \wedge (\imath_a *Q) \\
    { }^5\tau_a  &=&   (\imath_b Q) (\imath_c Q^{bc}) *e_a
                            - (\imath_a Q) (\imath_c Q^{bc}) *e_b
                            - (\imath_b Q) (\imath_a Q^{bc}) *e_c \; .
 \ea
One can consult Ref.\cite{adak2006} for the details of variations.
Since ${ }^0\tau_a  =  {R^b}_c \wedge *{e_{ab}}^c = 0$ and $D {
}^0{\Sigma_a}^b = D^2 *{e_a}^b \sim {R_a}^b =0 $ we drop the
Einstein-Hilbert term: $k_0 =0$. The case that $k_0 \neq 0$ and
others$=0$ was discussed in the previous subsection.

 \subsection{Spherical symmetric solution to the model}

Under the configuration (\ref{coframe1})-(\ref{connect1}) the only
nontrivial field equation comes from the trace of
(\ref{fieldeqn}):
 \ba
     d {\Sigma^a}_a + e^a \wedge \tau_a =0 \; . \label{65}
 \ea
Symmetric and antisymmetric parts of the field equation
(\ref{fieldeqn}) give trivially zero.  From (\ref{65}) we obtain
 \ba
     - \frac{\ell_1}{g} \left( \frac{f'}{fg} \right)'
     - \frac{\ell_2}{g} \left( \frac{1-g}{rg} \right)'
     + \ell_3 \left( \frac{f'}{fg} \right)^2
     + \ell_4 \left( \frac{f'}{fg} \right) \left( \frac{1-g}{rg} \right)
     - \frac{2\ell_1}{rg} \left( \frac{f'}{fg} \right) \nonumber \\
     - \frac{2\ell_2}{rg} \left( \frac{1-g}{rg} \right)
     + \ell_5 \left( \frac{1-g}{rg} \right)^2 = 0 \label{difeqn}
 \ea
where
 \ba
 \ell_1 &=& 2c_1 + 2c_2 + 8c_4 + c_5 ,\\
 \ell_2 &=& 6c_1 + 4c_2 + 2c_3 + 24c_4 + 7c_5 ,\\
 \ell_3 &=& - 6c_4 - c_5 ,\\
 \ell_4 &=& -6c_1 - 4c_2 - 2c_3 - 12c_4 - 5c_5 ,\\
 \ell_5 &=& 6c_1 + 4c_2 + 2c_3 + 18c_4 + 6c_5 \; .
 \ea
Mathematically, this equation has infinitely many solutions
because there are two functions and only one equation. We give two
classes of solutions. At first, let $f'/fg = u$, then
(\ref{difeqn}) takes the form, if $\ell_3 \neq 0$,
 \ba
     -\left( \ell_1 u + \ell_2 \frac{1-g}{rg} \right)'
     -\frac{2}{r} \left( \ell_1 u + \ell_2 \frac{1-g}{rg} \right)
     + \frac{\ell_3 g}{\ell_1^2} \left( \ell_1^2 u^2 + \frac{\ell_1^2 \ell_4}{\ell_3} u \frac{1-g}{rg}
     + \frac{\ell_1^2 \ell_5}{\ell_3}  \frac{(1-g)^2}{r^2g^2}
     \right) =0 \; . \label{difeqn1}
 \ea
When let $\ell_4 = 2\ell_2 \ell_3 / \ell_1$ and $\ell_5 = 2\ell_3
\ell_2^2 / \ell_1^2$ if we define
 \ba
     z= \ell_1 u + \ell_2 \frac{1-g}{rg}
 \ea
the equation becomes
 \ba
     - \frac{(r^2z)'}{(r^2z)^2} + \frac{\ell_3}{\ell_1^2}
     \frac{g}{r^2}=0 \; .
 \ea
This means that given $g$ we obtain $f$. At the second class;
$\ell_3 =0 \; , \ell_4 = \alpha \ell_1 \; , \ell_5 = \alpha
\ell_2$ with $\alpha \neq 0$ parameter, (\ref{difeqn}) turns out
to be
 \ba
 - \frac{(r^2z)'}{r^2z} + \alpha \frac{1-g}{r}=0 \; .
 \ea
Again to specify $g$ yields $f$. Physically, however, due to the
observational success of the Schwarzschild solution of general
relativity, we investigate solutions with $ g=1/f $. Then
(\ref{difeqn}) becomes
 \ba
  - \ell_1 ff'' + \ell_3 (f')^2 - (2\ell_1 + \ell_2 - \ell_4) \frac{ff'}{r}
     + (\ell_5 - \ell_2) \frac{f^2}{r^2} - \ell_4 \frac{f'}{r} + ( \ell_2 - 2\ell_5) \frac{f}{r^2}
     + \ell_5 \frac{1}{r^2}= 0 \; . \label{srn-equation}
 \ea
We can not find an analytical exact solution of this nonlinear
second order differential equation. Therefore, we treat the linear
sector of the equation
 \ba
     r^2 \left(f^2\right)'' + (2+\frac{\ell_2}{\ell_1}) r
     \left(f^2\right)' + \frac{\ell_2}{\ell_1} f^2 = \frac{\ell_2}{\ell_1}
 \ea
by choosing our parameters as follows;
 \ba
     \ell_3 = -\ell_1  \quad , \quad  \ell_4 =0  \quad , \quad \ell_5 = \frac{\ell_2}{2} \; .
 \ea
Here some special cases deserve attention.

\begin{enumerate}
    \item For $\ell_2 =\ell_1 $, we obtain the solution
                  \ba
               f^2 = 1 + \frac{C_1}{r} + D_1 \frac{\ln{r}}{r}
                 \ea
            which is asymptotically flat; $\lim_{r \rightarrow \infty} f = 1$.
            Here $C_1$ and $D_1$ are integration constants.
    \item For $\ell_2 \neq \ell_1 $, the solution is found as
                \ba
              f^2 = 1 + \frac{C_2}{r} + \frac{D_2}{r^{\ell_2 / \ell_1}}
                \ea
           where $C_2$ and $D_2$ are integration constants.
\begin{enumerate}
    \item  For $\ell_2 =0$, we obtain a Schwarzschild-type solution with
           $D_2=0$ for asymptotically flatness and we identify the other
           constant with a spherically symmetric mass centered at the origin; $C_2 = -2M$.
    \item  For $\ell_2 =-2\ell_1$, we obtain a Schwarzschild-de Sitter-type solution.
           We again identify $C_2$ with mass $C_2 = -2M$ and $D_2$ with
           cosmological constant $D_2=-\frac{1}{3}\Lambda$. The $\Lambda$ term corresponds to a
           repulsive central force of magnitude $\frac{1}{3}\Lambda r$, which
           is independent of the central mass.
    \item  For $\ell_2 = 2\ell_1$, we obtain a Reissner-Nordstr\"{o}m-type solution.
           We again identify $C_2$ with mass $C_2 = -2M$ while $D_2$ with a new
           kind of gravitational charge. We hope that besides ordinary matter
           that interacts gravitationally through its mass, the dark matter
           in the Universe may interact gravitationally through both its mass
           and this new gravitational charge.
\end{enumerate}
\end{enumerate}

\section{Conclusion}

In this paper we investigated the symmetric teleparallel gravity.
After giving the irreducible decompositions of non-metricity under
the Lorentz group we identified STPG theories and gave an approach
for the generic solution: {\it natural or inertial gauge}. Then we
obtained symmetric teleparallel equivalence of the
Einstein-Hilbert Lagrangian and found a spherically symmetric
static solution to Einstein's equation in STP geometry. We
analyzed the singularity structure of the space-time according to
that solution. The singularities need to be clarified in the
non-Riemannian space-times. Finally, we studied the Lagrange
formulation of the general STPG by considering a 5-parameter
symmetric teleparallel Lagrangian without {\it a priori}
restricting the coupling constants $c_1, c_2, c_3, c_4, c_5$. We
obtained sets of solutions in which the Schwarzschild-type,
Schwarzschild-de Sitter-type and Reissner-Nordstr\"{o}m-type
solutions are some physically interesting ones. Consequently, we
suggest that in addition ordinary matter that interacts
gravitationally through its mass, the dark matter in the Universe
may interact gravitationally through both its mass and a new kind
of gravitational charge \cite{tucker1998}-\cite{dereli2000}. The
latter coupling is analogous to the coupling of electric charge to
electromagnetic field where the analogue of the Maxwell field is
the non-metricity field strength. That is, such "charges" may
provide a source for the non-metricity. The novel gravitational
interactions may have a significant influence on the structure of
black holes. For example, we may speculate that this unknown
gravitational charge may have repulsive nature.

\section*{Acknowledgment}

This paper was presented at the 5th Workshop on Quantization,
Dualities and Integrable Systems, Pamukkale University, 23-28
January 2006, Denizli, Turkey. The author is grateful to  Prof.
Dr. Metin G\"{U}RSES  for his fruitful comments.

\end{document}